\documentclass[journal]{IEEEtran}

\usepackage{cite}
\usepackage{graphicx}
\usepackage{url}
\usepackage{amsmath,amssymb}
\usepackage{booktabs}
\usepackage{array}
\usepackage{enumitem}
\usepackage{xcolor}
\usepackage{tikz}
\usepackage{siunitx}

\usetikzlibrary{shapes.geometric, arrows}

\tikzstyle{startstop} = [rectangle, rounded corners, minimum width=3cm, minimum height=1cm,text centered, draw=black, fill=red!20]
\tikzstyle{process}   = [rectangle, minimum width=3cm, minimum height=1cm, text centered, draw=black, fill=blue!20]
\tikzstyle{decision}  = [diamond, minimum width=3cm, minimum height=1cm, text centered, draw=black, fill=green!20]
\tikzstyle{arrow}     = [thick,->,>=stealth]

\allowdisplaybreaks

\begin{document}

\title{VAST: Vascular Flow Analysis and Segmentation for Intracranial 4D Flow MRI}

\author{Abhishek~Singh, Vitaliy~L.~Rayz, and Pavlos~P.~Vlachos% 
\thanks{A. Singh is with the School of Mechanical Engineering, Purdue University, West Lafayette, IN 47907 USA (e-mail: sing1062@purdue.edu).}%
\thanks{V. L. Rayz and P. P. Vlachos are with the School of Mechanical Engineering and Weldon School of Biomedical Engineering, Purdue University, West Lafayette, IN 47907 USA (e-mail: vrayz@purdue.edu; pvlachos@purdue.edu).}%
\thanks{This work was supported by the National Institutes of Health (NIH), National Heart, Lung, and Blood Institute (NHLBI) under Grant R01 HL115267.}%
} 

\maketitle

\begin{abstract}
Four-dimensional (4D) Flow MRI can noninvasively measure cerebrovascular hemodynamics but remains underused clinically because current workflows rely on manual vessel segmentation and yield velocity fields sensitive to noise, artifacts, and phase aliasing. We present VAST (Vascular Flow Analysis and Segmentation), an automated, unsupervised pipeline for intracranial 4D Flow MRI that couples vessel segmentation with physics-informed velocity reconstruction. VAST derives vessel masks directly from complex 4D Flow data by iteratively fusing magnitude- and phase-based background statistics. It then reconstructs velocities via continuity-constrained phase unwrapping, outlier correction, and low-rank denoising to reduce noise and aliasing while promoting mass-consistent flow fields, with processing completing in minutes per case on a standard CPU. We validate VAST on synthetic data from an internal carotid artery aneurysm model across SNR = 2--20 and severe phase wrapping (up to five-fold), on in vitro Poiseuille flow, and on an in vivo internal carotid aneurysm dataset. In synthetic benchmarks, VAST maintains near quarter-voxel surface accuracy and reduces velocity root-mean-square error by up to fourfold under the most degraded conditions. In vitro, it segments the channel within approximately half a voxel of expert annotations and reduces velocity error by 39\% (unwrapped) and 77\% (aliased). In vivo, VAST closely matches expert time-of-flight masks and lowers divergence residuals by about 30\%, indicating a more self-consistent intracranial flow field. By automating processing and enforcing basic flow physics, VAST helps move intracranial 4D Flow MRI toward routine quantitative use in cerebrovascular assessment.
\end{abstract}


\section{Introduction}

Cerebrovascular disease is a major cause of stroke, long-term disability, and death worldwide \cite{feigin2021global, cdc2023stroke}. Intracranial aneurysms alone affect nearly 1 in 50 individuals in the United States and account for tens of thousands of ruptures each year \cite{bafound2023aneurysm, unruptured1998, malhotra2017growth}. Current diagnostic assessment relies on structural imaging with computed tomography (CT), CT angiography (CTA), and magnetic resonance imaging (MRI), which are optimized to detect hemorrhage, vessel caliber, and aneurysm morphology \cite{wintermark2013imaging, chen2018meta}. These studies, however, do not characterize the blood-flow environment. Abnormal flow patterns—such as recirculation zones, elevated shear, and disturbed wall shear stress (WSS)—are central drivers of aneurysm growth, thrombus formation, and vascular remodeling \cite{yuan2025hemo, chiu2011disturbed, meng2014wss}.

Four-dimensional (4D) Flow MRI can measure volumetric, time-resolved blood velocity in vivo \cite{markl20124dflow} and enables estimation of intracranial flow patterns, WSS, and pressure surrogates \cite{wahlin2022biomarkers, rivera20164d, pressure, wss}. However, routine cerebrovascular use remains limited because reliable, reproducible hemodynamic quantification is difficult under typical acquisition constraints. Spatial and temporal resolution limit depiction of small vessels and transient features \cite{baiyan20214d}, cardiac averaging and scan time reduce sensitivity to subtle dynamics \cite{callmer2025deep}, and noise, intravoxel dephasing, aliasing, and partial-volume effects degrade velocity accuracy—especially near the vessel wall \cite{bissell20234dconsensus, biasmodelling, unwrapping, rispoli_synMRI}. As a result, extracting robust biomarkers hinges on two coupled problems: dependable vessel segmentation and velocity reconstruction that is internally consistent with basic flow physics.

Segmentation remains a major workflow bottleneck. Clinically, experts often delineate intracranial vessels on high-resolution MR angiography (MRA) and transfer masks to the 4D Flow volume \cite{wintermark2013imaging}, a process that is labor-intensive, operator-dependent, and sensitive to inter-scan mismatch. Automatic segmentation methods operating directly on 4D Flow MRI either rely primarily on magnitude contrast \cite{4dcls1, pcd}—which can fail in slow-flow regions and vortex cores—or require supervised training data that may not generalize across anatomically diverse patients and acquisition protocols \cite{pcd-nn}. Phase-based approaches, including our earlier Standardized Difference of Means (SDM) method \cite{sdm}, improve robustness in principle but still depend on hand-tuned hyperparameters and acquisition-specific calibration.

Velocity reconstruction and post-processing present complementary challenges. Prospective deep-learning and compressed-sensing approaches can reduce noise and accelerate acquisition \cite{vishnevskiy2020deep, nath22flowrau-net, jacobsself}, but are typically supervised, do not jointly solve segmentation, and generally do not enforce flow-physics constraints such as mass conservation. Related label-free formulations instead couple segmentation and velocity inference via a shared probabilistic measurement model for 4D Flow MRI~\cite{hans2025smurf}. Retrospective strategies—including super-resolution networks and Physics-Informed Neural Networks (PINNs)—can incorporate physics during enhancement \cite{4dflownet, msr, 4dflownetphysics, neal_nn, deepphysicscardiac, suk2024physics, pinns_Re_stress}, but they also commonly rely on paired training data (e.g., CFD or carefully curated experiments) and are often demonstrated on limited anatomies with predefined segmentations. Establishing general performance across cerebrovascular geometries, flow environments, and routine acquisition settings therefore remains an open need.

To address these limitations, we introduce \emph{VAST} (Vascular Flow Analysis and Segmentation), a fully unsupervised workflow for intracranial 4D Flow MRI that derives vessel segmentations directly from the complex data and subsequently corrects and denoises the phase measurements to improve robustness to noise and phase aliasing. Our contributions are:
(1) an automatic segmentation approach that derives intracranial vessel masks directly from complex 4D Flow MRI without acquisition-specific threshold tuning;
(2) a continuity-constrained reconstruction procedure that improves velocity self-consistency in the presence of noise and phase wrapping; and
(3) validation on synthetic aneurysm benchmarks, an in vitro Poiseuille-flow experiment, and an in vivo intracranial aneurysm scan using geometric segmentation metrics, velocity error metrics where ground truth is available, and divergence residuals as an in vivo consistency measure.

The remainder of the paper is organized as follows. Section~\ref{sec:methodology} summarizes the workflow and datasets. Sections~\ref{sec:synthetic}--\ref{sec:invivo} present results in synthetic, in vitro, and in vivo settings. Section~\ref{sec:computational} reports computational performance, and Section~\ref{sec:conclusion} discusses implications and future extensions.

\section{Methodology}\label{sec:methodology}

\begin{figure}[!t]
    \centering
    \includegraphics[width=\columnwidth]{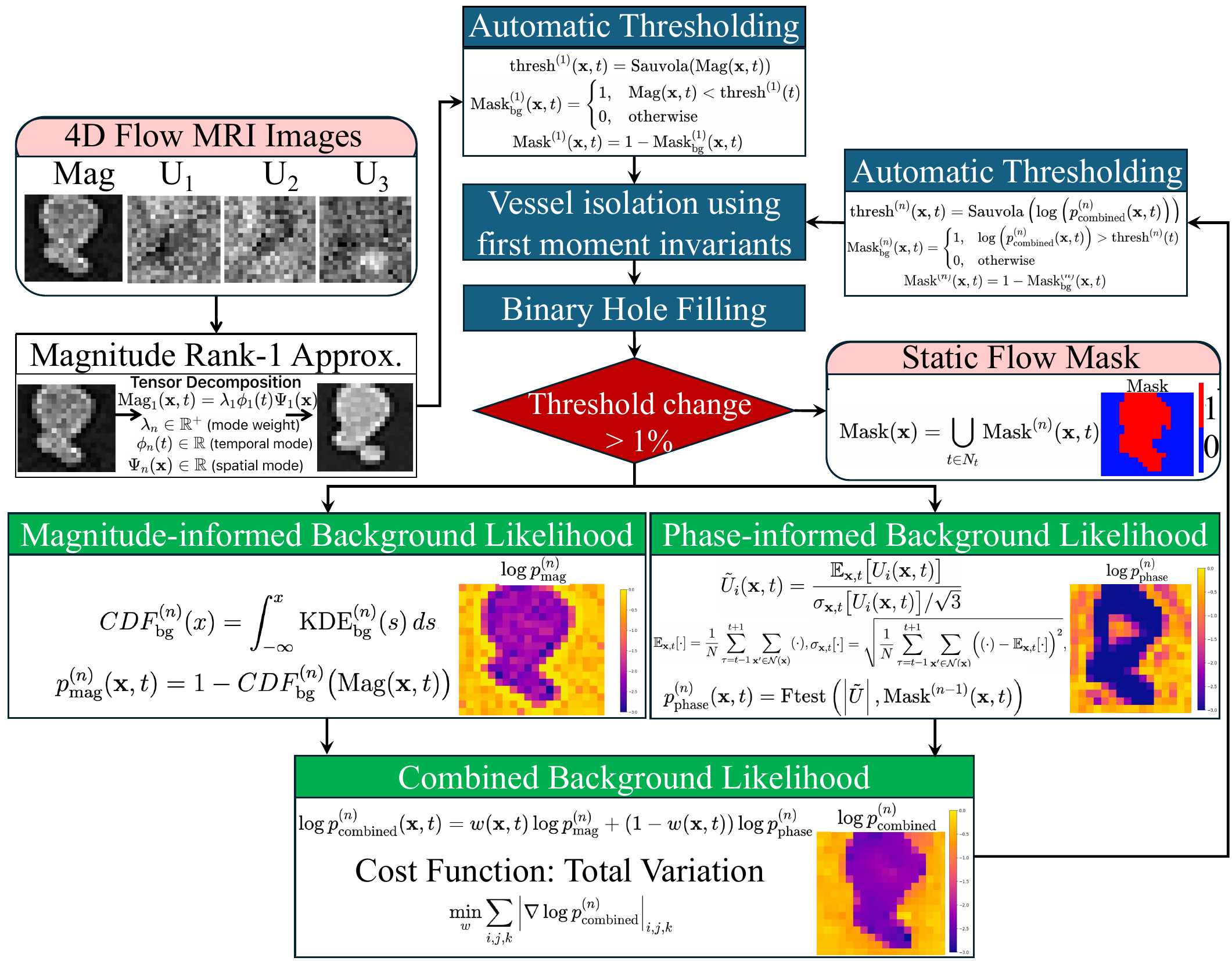}
    \caption{VAST segmentation workflow. Magnitude images are denoised via low-rank tensor decomposition to initialize a background mask using adaptive thresholding. At each iteration, background statistics are updated to compute magnitude-based and phase-based (SDM) background likelihoods. These likelihoods are fused into a combined background likelihood map using a spatially regularized weighting field (total-variation regularization) to preserve vessel boundaries while promoting spatial coherence. Adaptive thresholding yields an updated flow mask, which is refined by vessel isolation and hole filling. The procedure repeats until convergence (Appendix \ref{app:likelihood_fusion}).}
    \label{fig:method_segmentation}
\end{figure}

\begin{figure}[!t]
    \centering
    \includegraphics[width=\columnwidth]{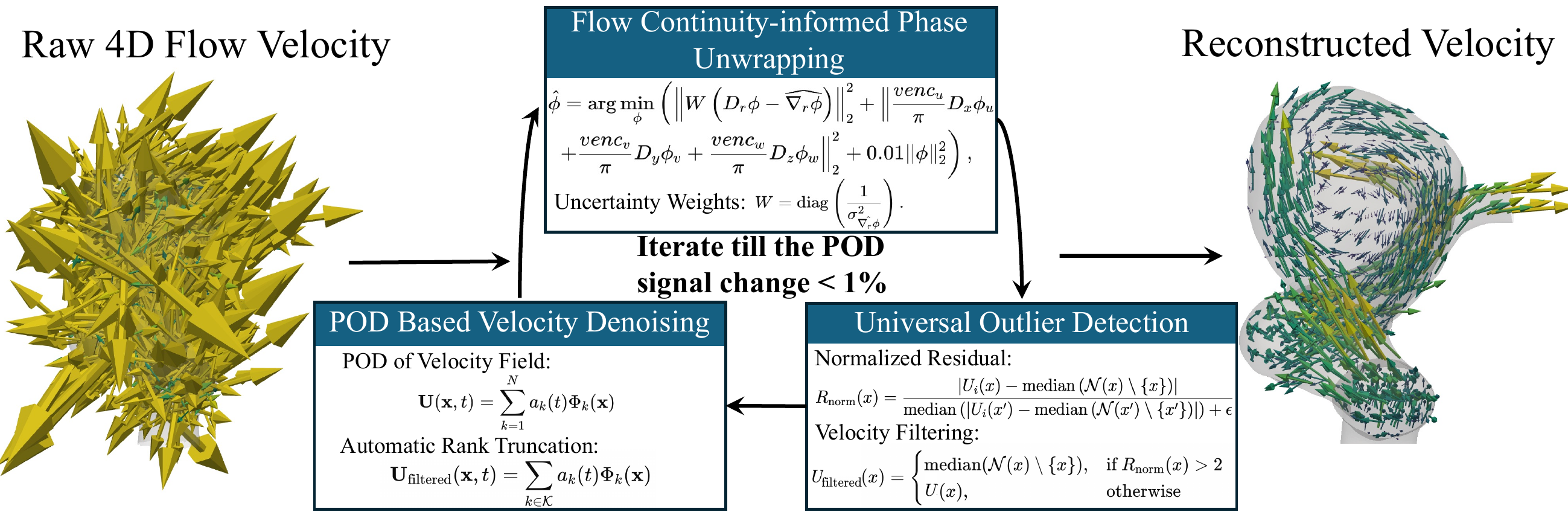}
    \caption{VAST velocity reconstruction workflow. Within the segmented flow volume, phase images are unwrapped under a mass-continuity constraint to obtain continuity-consistent velocity fields. Localized spurious voxels are corrected using universal outlier detection, and the resulting 4D velocity field is denoised in a low-rank POD basis to exploit temporal coherence. Phase unwrapping, outlier correction, and POD denoising are applied iteratively until the reconstruction stabilizes under an energy-based criterion (Appendix \ref{app:iteration}).}
    \label{fig:method_reconstruction}
\end{figure}

Figure~\ref{fig:method_segmentation} summarizes the segmentation workflow used to derive an intracranial flow mask directly from the 4D Flow magnitude and phase data, and Figure~\ref{fig:method_reconstruction} summarizes the subsequent velocity reconstruction workflow performed within this mask. Appendix~A provides implementation details and parameter settings.

\subsection{Segmentation}

Segmentation estimates, at each cardiac phase, the probability that a voxel belongs to background versus flowing blood and then aggregates the phase-wise masks into a static intracranial vessel mask for reconstruction. Magnitude images are first denoised using low-rank tensor decomposition (Appendix \ref{app:magnitude_denoising}), and an initial background--flow partition is obtained using adaptive Sauvola thresholding (Appendix \ref{app:initial_background})~\cite{sauvola}.

VAST refines the initial mask via iterative background likelihood estimation from both magnitude and phase. Background magnitude statistics are updated from the current background voxels to compute a magnitude-informed background likelihood (Appendix \ref{app:mag_likelihood}). In parallel, a phase-informed likelihood is computed using an extension of the Standardized Difference of Means (SDM) framework~\cite{sdm}, which treats background phase as noise and identifies voxels whose spatiotemporal velocity statistics deviate from this model (Appendix \ref{app:phase_likelihood}).

The two likelihoods are fused into a single background probability field using a spatially regularized, automatically determined weighting that promotes coherence while preserving vessel boundaries (Appendix \ref{app:likelihood_fusion}). Thresholding yields an updated mask, followed by morphological cleanup and vessel isolation adapted from SDM~\cite{sdm}. Iteration continues until the segmentation stabilizes (Appendix \ref{app:likelihood_fusion}).

In brief, VAST forms a convex combination of the magnitude- and phase-informed log-likelihoods, with a spatially varying weight field that is automatically selected via total-variation regularization. This fusion adapts on a per-case basis: it relies more heavily on magnitude contrast where the magnitude is reliable, and shifts toward phase statistics in slow-flow or low-contrast regions where magnitude contrast can be ambiguous.

Because intracranial vessel-wall motion is negligible for the applications considered here, a static mask is formed by taking the union of the phase-wise masks across the cardiac cycle. This static mask is used for reconstruction and for comparison with PCD and SDM~\cite{pcd,sdm}.

\subsection{Flow reconstruction}

Within the segmented flow domain (zero velocity enforced outside the mask), reconstruction corrects phase wrapping, suppresses noise, and removes localized artifacts. Phase aliasing is addressed by unwrapping the phase images under a mass-continuity constraint. When the velocity-encoding parameter (\(v_{\text{enc}}\)) is lower than the true peak velocity, the measured phase wraps into \([-\pi,\pi]\). We recover the unwrapped phase by solving a constrained least-squares problem that (i) enforces consistency with wrapped phase gradients and (ii) penalizes divergence of the implied velocity field, following our previously proposed framework~\cite{unwrapping}. This yields a continuity-conditioned phase \(\hat{\phi}\) for each encoding direction (Appendix \ref{app:unwrapping}), from which velocities are obtained via the standard phase--velocity relationship.

Even after continuity-constrained unwrapping, localized outliers can persist, particularly near the vessel wall where partial-volume effects and background contamination are most pronounced. Universal Outlier Detection (UOD)~\cite{uod} is therefore applied to identify voxels whose velocity components deviate strongly from local neighborhood statistics, replacing flagged values with neighborhood medians (Appendix \ref{app:uod}).

Finally, temporal coherence is exploited by denoising the 4D velocity field using Proper Orthogonal Decomposition (POD). The field is decomposed into spatial modes and temporal coefficients; modes are ranked by an information-content measure based on spectral entropy, and coherent flow modes are separated from noise-dominated modes using density-based clustering (Appendix \ref{app:pod}). Unwrapping, outlier correction, and POD denoising are applied in an outer iteration until the reconstruction stabilizes under the POD-energy criterion defined in Appendix \ref{app:iteration}.

Even in unaliased acquisitions, the continuity-constrained unwrapping step provides a global phase denoiser by reconciling wrapped gradient observations with a divergence-penalized phase field. The subsequent UOD and POD stages target complementary artifact classes---localized spurious voxels and temporally incoherent fluctuations, respectively---and the full outer loop typically converges in a small number of iterations (\S\ref{sec:computational}).

\subsection{Datasets}

\subsubsection{Synthetic 4D Flow MRI in an internal carotid artery aneurysm}

To enable controlled benchmarking with known ground truth, we generated synthetic 4D Flow MRI from a patient-specific internal carotid artery (ICA) aneurysm model using a time-resolved CFD simulation~\cite{brindise_multimodality}. The CFD solution provides a three-component velocity field over the cardiac cycle, which we mapped onto a 4D Flow sampling grid and used to emulate a four-point phase-contrast acquisition by assigning complex-valued signals for each encoding direction at a selected \(v_{\text{enc}}\).

To approximate image formation, a three-dimensional point-spread function was applied to mimic intravoxel averaging and complex Gaussian noise was added to obtain predefined SNR levels. We varied SNR from moderate to very low and \(v_{\text{enc}}\) from fully resolving to strongly aliased settings, producing cases with known reference geometry and velocities for evaluation. Full details are provided in Appendix \ref{app:synthetic}.

\subsubsection{Experimental Poiseuille flow}

To evaluate VAST on a real 4D Flow MRI acquisition where both the velocity field and geometry admit independent references, we performed a straight-tube Poiseuille-flow experiment on a GE Discovery MR750 3~T scanner at the Purdue MRI Facility. The expected laminar profile provides an analytical benchmark for reconstruction, while a high-resolution time-of-flight (TOF) scan provides an independent geometric reference for segmentation. A Compuflow~1000~MR piston pump drove steady flow (\(\approx 30\,\mathrm{cm/s}\)) of a 60:40 water--glycerol mixture (density \(1110\,\mathrm{kg/m^3}\), viscosity \(0.0035\,\mathrm{Pa\cdot s}\)) through a quarter-inch PDMS channel. The 4D Flow acquisition used a native voxel size of \(1.0\times1.0\times1.0\,\mathrm{mm^3}\) with 2$\times$ ZIP interpolation to \(0.5\times0.5\times0.5\,\mathrm{mm^3}\), nine cardiac phases, \(\mathrm{TR}=9.15\,\mathrm{ms}\), \(\mathrm{TE}=3.50\,\mathrm{ms}\), and flip angle \(8^\circ\). Two scans were acquired: an unaliased dataset at \(v_{\text{enc}}=70\,\mathrm{cm/s}\) and an aliased dataset at \(v_{\text{enc}}=50\,\mathrm{cm/s}\), each with an acquisition time of approximately four minutes.

A TOF angiogram was acquired at \(0.5078\times0.5078\times1.0\,\mathrm{mm^3}\) resolution (\(\mathrm{TR}=3.552\,\mathrm{ms}\), \(\mathrm{TE}=1.252\,\mathrm{ms}\), flip angle \(8^\circ\)). The channel lumen was semi-automatically segmented in ITK-SNAP and rigidly registered to the VAST-derived 4D Flow segmentation using Open3D for quantitative comparison~\cite{itksnap, open3d}.

\subsubsection{In vivo internal carotid artery aneurysm imaging}\label{sec:invivo_data}

To demonstrate in vivo applicability, we analyzed 4D Flow MRI data from a patient with an ICA aneurysm scanned on a Siemens Skyra 3~T system at Northwestern Memorial Hospital~\cite{brindise_multimodality}. TOF angiography was acquired at \(0.4\times0.4\times0.6\,\mathrm{mm^3}\) resolution, followed by 4D Flow MRI with \(\mathrm{TE/TR}=2.997/6.4\,\mathrm{ms}\), flip angle \(15^\circ\), \(v_{\text{enc}}=80\,\mathrm{cm/s}\), temporal resolution \(44.8\,\mathrm{ms}\), and spatial resolution \(1.09\times1.09\times1.30\,\mathrm{mm^3}\). TOF-derived lumen segmentations were rigidly registered to the VAST segmentation for geometric comparison, consistent with the in vitro validation procedure.

\section{Results and Discussion}

We evaluate VAST on synthetic, in vitro, and in vivo datasets. Segmentation performance is compared against PCD and SDM \cite{pcd, sdm}, and velocity reconstruction is evaluated against raw 4D Flow MRI measurements using CFD- or experiment-derived reference data.

\subsection{Synthetic benchmarking}\label{sec:synthetic}

We first evaluate VAST on synthetic 4D Flow MRI generated from a patient-specific internal carotid artery (ICA) aneurysm model, where CFD provides ground-truth geometry and velocities. The flow contains a systolic jet, intra-aneurysmal recirculation, and branching outflow into the posterior communicating artery (PComA), anterior cerebral artery (ACA), and middle cerebral artery (MCA) (Fig.~\ref{fig:synthetic_vel_viz}). We vary SNR and velocity-encoding (\(v_{\text{enc}}\)) to isolate sensitivity to noise and phase wrapping.

\subsubsection{Segmentation}

\begin{figure}[!t]
    \centering
    \includegraphics[width=\columnwidth]{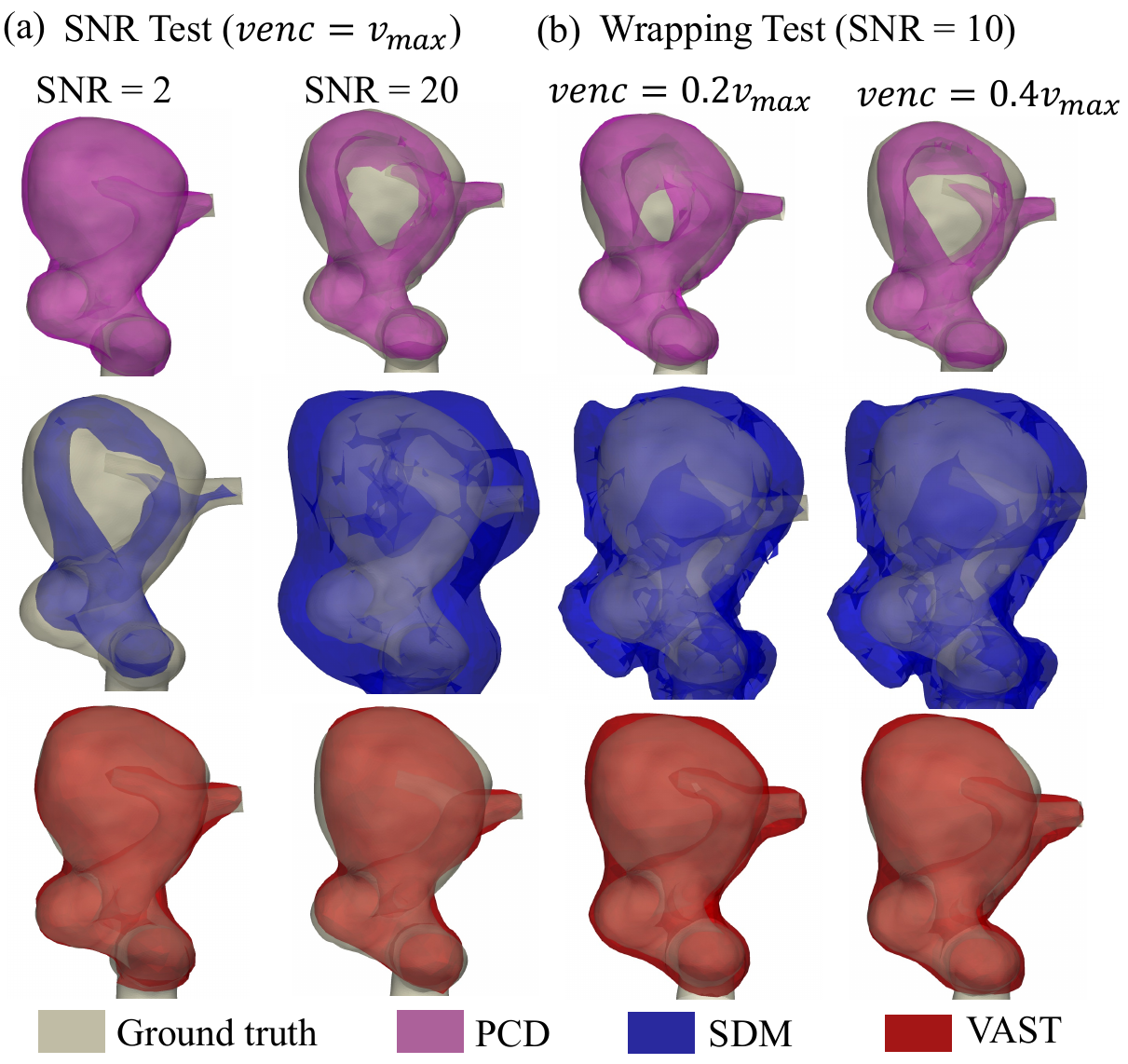}
    \caption{Synthetic ICA aneurysm segmentation. PCD, SDM, and VAST surfaces are overlaid on the ground-truth geometry for (a) two noise levels (\(SNR=20\) and \(SNR=2\)) and (b) two velocity-encoding settings (\(v_{\text{enc}}=0.4\,v_{\max}\) and \(v_{\text{enc}}=0.2\,v_{\max}\)).}
    \label{fig:synthetic_seg_viz}
\end{figure}

\begin{figure}[!t]
    \centering
    \includegraphics[width=\columnwidth]{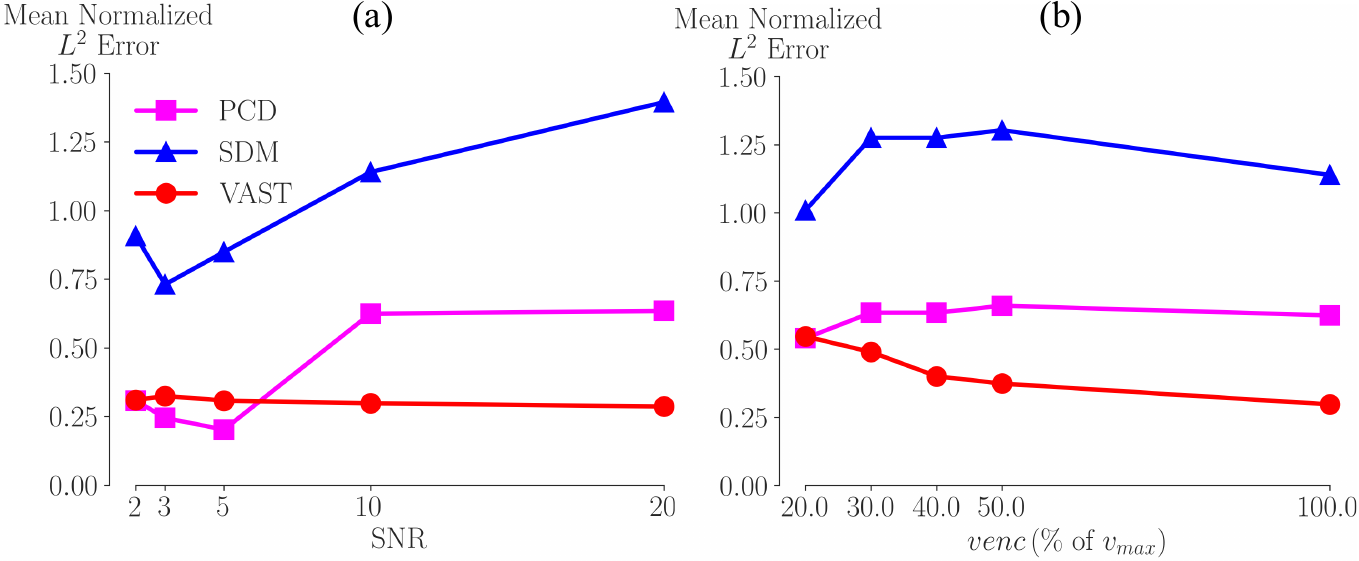}
    \caption{Surface-based segmentation error on the synthetic ICA aneurysm. Mean Euclidean surface distance between each automated segmentation (PCD, SDM, VAST) and the ground-truth mesh across (a) SNR and (b) \(v_{\text{enc}}\). Distances are normalized by the smallest voxel dimension.}
    \label{fig:synthetic_seg_quant}
\end{figure}

\begin{figure}[!t]
    \centering
    \includegraphics[width=\columnwidth]{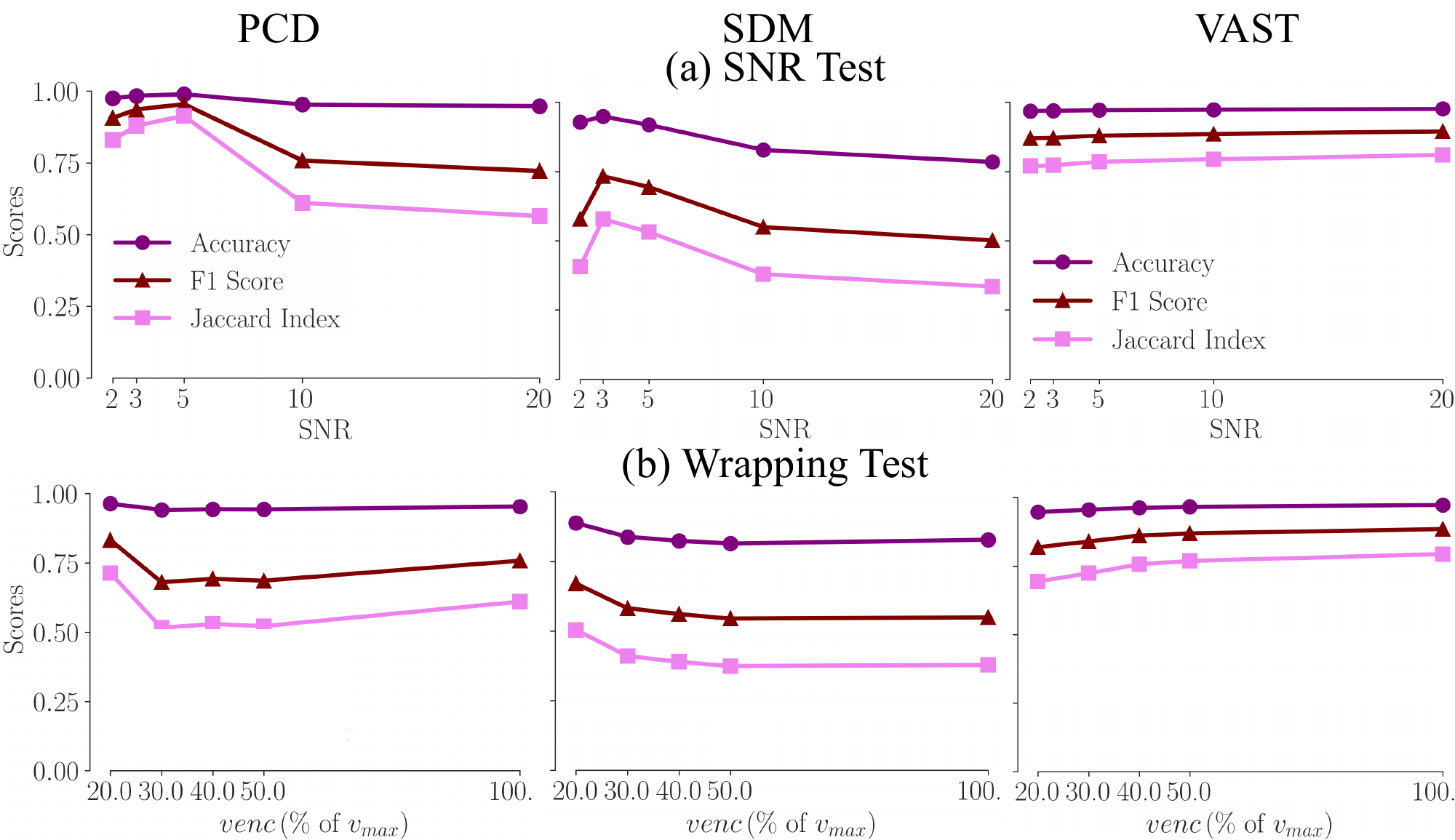}
    \caption{Volumetric segmentation metrics on the synthetic aneurysm dataset. Accuracy, F1-score, and Jaccard similarity (definitions in Table~\ref{tab:seg_scores_definition}) are shown for PCD, SDM, and VAST across (a) noise levels (SNR sweep) and (b) velocity-encoding settings (\(v_{\text{enc}}\) sweep).}
    \label{fig:synthetic_seg_quant_volume}
\end{figure}

Across both sweeps, VAST achieves the most consistent agreement with the ground-truth lumen, attaining the lowest surface error and the highest overlap metrics relative to PCD and SDM (Figs.~\ref{fig:synthetic_seg_quant}--\ref{fig:synthetic_seg_quant_volume}). Representative surface overlays are shown in Fig.~\ref{fig:synthetic_seg_viz}.

Qualitatively, PCD exhibits reduced fidelity at higher SNR, where its heavy reliance on velocity-magnitude/complex-difference contrast becomes more dominant and tends to under-segment slow-flow regions such as vortex cores and near-wall flow. At lower SNR, increased noise reduces the relative dominance of the velocity magnitude-driven term in the PCD criterion, which can mitigate this slow-flow erosion. SDM exhibits reduced stability in this synthetic setting, consistent with reliance on fixed statistical thresholds and smoothing choices tuned for in vivo data. In contrast, VAST preserves the aneurysm sac and primary branches more reliably across both noise- and wrapping-dominated regimes; residual discrepancies are largely confined to smaller distal segments at the most severe corruptions.

Surface-distance summaries (Fig.~\ref{fig:synthetic_seg_quant}) quantify these trends. VAST remains sub-voxel on average across the SNR sweep and shows a smaller degradation than PCD at \(SNR \ge 5\). Under increased wrapping (lower \(v_{\text{enc}}\)), all methods lose accuracy, but VAST maintains the lowest mean surface error over the full \(v_{\text{enc}}\) range.

Volumetric metrics follow the same ordering (Fig.~\ref{fig:synthetic_seg_quant_volume}). Because the volume is background-dominated, accuracy is less discriminative than overlap metrics; we therefore emphasize F1-score and Jaccard similarity. VAST maintains overlap scores on the order of \(\sim 0.7\) or higher across both sweeps, while PCD drops sharply at higher SNRs due to reduced recall and SDM is generally lower across conditions.

\begin{table}[htbp]
    \centering
    \caption{Volumetric segmentation metrics used for quantitative evaluation.}
    \label{tab:seg_scores_definition}
    \renewcommand{\arraystretch}{1.3}
    \setlength{\tabcolsep}{5pt}
    \small
    \begin{tabular}{c c}
    \toprule
    \textbf{Metric} & \textbf{Formula} \\
    \midrule
    Accuracy & $\dfrac{|T \cap P| + |\neg T \cap \neg P|}{|T \cup P| + |\neg T \cup \neg P|}$ \\[1ex]
    Precision & $\dfrac{|T \cap P|}{|P|}$ \\[1ex]
    Recall & $\dfrac{|T \cap P|}{|T|}$ \\[1ex]
    F1-score & $\dfrac{2|T \cap P|}{2|T \cap P| + |T \setminus P| + |P \setminus T|}$ \\[1ex]
    Dice coefficient & $\dfrac{2|T \cap P|}{|T| + |P|}$ \\[1ex]
    Jaccard similarity & $\dfrac{|T \cap P|}{|T \cup P|}$ \\
    \bottomrule
    \multicolumn{2}{c}{
      \textbf{Notation:}
    }\\[0.5ex]
    \multicolumn{2}{c}{
      \begin{minipage}{0.8\linewidth}
        \small
        \begin{itemize}[leftmargin=*]
          \item $T$: voxels labeled as flow in the ground truth.
          \item $P$: voxels labeled as flow in the prediction.
          \item $\neg T$: voxels labeled as background in the ground truth.
          \item $\neg P$: voxels labeled as background in the prediction.
          \item $\cap$: intersection; $\cup$: union; $\setminus$: set difference; $|\cdot|$: cardinality.
        \end{itemize}
      \end{minipage}
    }\\
    \end{tabular}
\end{table}

\subsubsection{Flow reconstruction}

\begin{figure}[!t]
    \centering
    \includegraphics[width=\columnwidth]{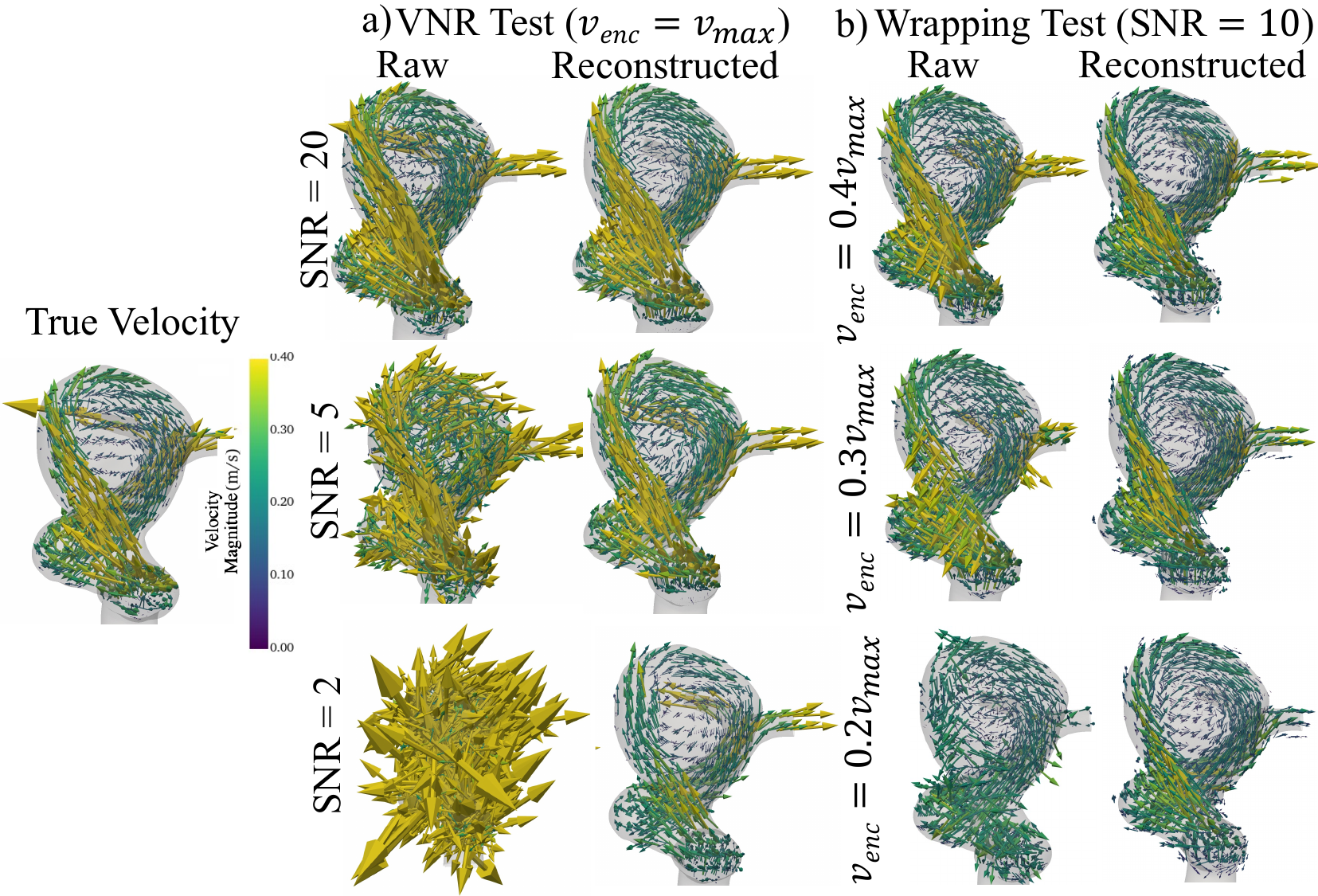}
    \caption{Synthetic ICA aneurysm: peak-systolic velocity vector fields for representative corruption settings. Results are shown for three noise levels (\(SNR = 20, 5, 2\); panel a) and three velocity-encoding settings (\(v_{\text{enc}} = 0.5, 0.3, 0.2\,v_{\max}\); panel b), comparing the CFD reference, raw 4D Flow measurements, and VAST-reconstructed velocities.}
    \label{fig:synthetic_vel_viz}
\end{figure}

We evaluate reconstruction by comparing raw 4D Flow measurements and VAST outputs to the CFD reference within the segmented domain. Representative peak-systolic vector fields (Fig.~\ref{fig:synthetic_vel_viz}) illustrate that at low SNR the raw field becomes noise-dominated, whereas VAST recovers coherent large-scale organization and suppresses high-frequency fluctuations. Under decreasing \(v_{\text{enc}}\), phase wrapping induces local direction reversals in the raw vectors; VAST substantially reduces these directional inconsistencies, although the most strongly wrapped regime shows some attenuation of peak speeds.

\begin{figure}[!t]
    \centering
    \includegraphics[width=\columnwidth]{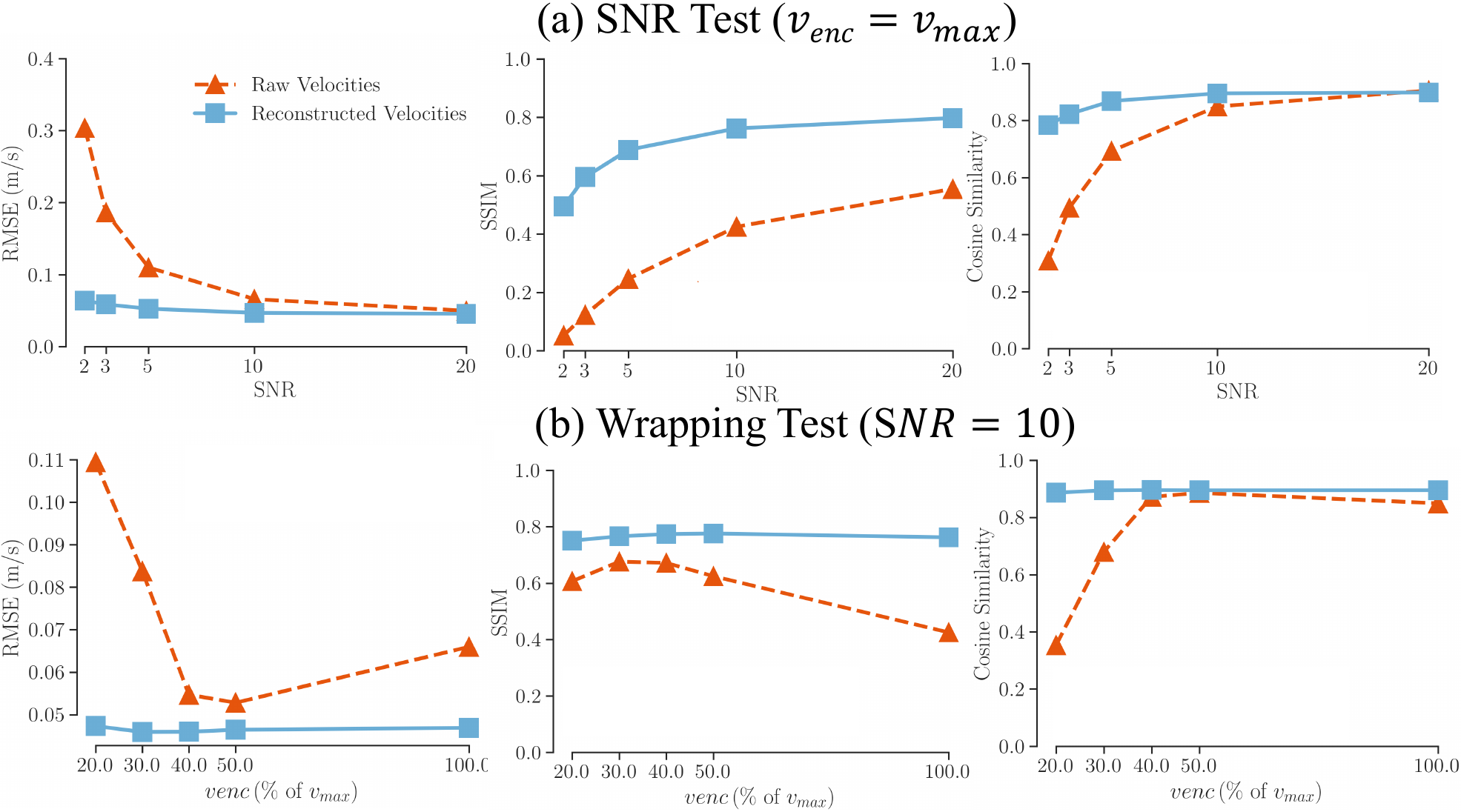}
    \caption{Synthetic aneurysm velocity reconstruction: global agreement metrics relative to the CFD reference. RMSE, SSIM, and cosine similarity are computed for raw and VAST-reconstructed velocity fields and reported across (a) the SNR sweep and (b) the \(v_{\text{enc}}\) sweep.}
    \label{fig:synthetic_recon_quant}
\end{figure}

Global agreement metrics (Fig.~\ref{fig:synthetic_recon_quant}) summarize performance. RMSE captures magnitude error, SSIM quantifies structural similarity (higher is better), and cosine similarity measures directional alignment with the reference field. Across the SNR sweep, the improvements are most pronounced in the noise-dominated regime: at the lowest SNR, VAST reduces RMSE by more than a factor of four and recovers substantial structural and directional agreement (SSIM increasing from near-zero to above \(\sim 0.5\), and cosine similarity increasing from \(\sim 0.3\) to \(\sim 0.8\)). These gains are consistent with the reconstruction stage suppressing high-frequency noise and enforcing global consistency through continuity-constrained unwrapping.

Across the \(v_{\text{enc}}\) sweep, raw errors initially decrease because lower \(v_{\text{enc}}\) reduces effective velocity noise at fixed SNR~\cite{vnr}, but then increase once wrapping becomes prevalent and local direction reversals appear. Over this transition, VAST improves all three metrics relative to raw and remains robust even at the most aliased setting, where performance is broadly comparable to the unwrapped regime despite modest attenuation of peak speeds.

\subsection{In vitro validation}\label{sec:experimental}

We next evaluate VAST on a controlled 4D Flow MRI experiment consisting of steady Poiseuille flow through a straight PDMS channel with circular cross-section. This setting provides two independent references: (i) the analytical parabolic axial-velocity profile for reconstruction and (ii) a high-resolution TOF-derived channel geometry for segmentation. We acquire two 4D Flow datasets: an unaliased scan (\(v_{\text{enc}} = 70~\mathrm{cm/s}\)) and an aliased scan (\(v_{\text{enc}} = 50~\mathrm{cm/s}\)).

\subsubsection{Segmentation}

\begin{figure}[!t]
    \centering
    \includegraphics[width=\columnwidth]{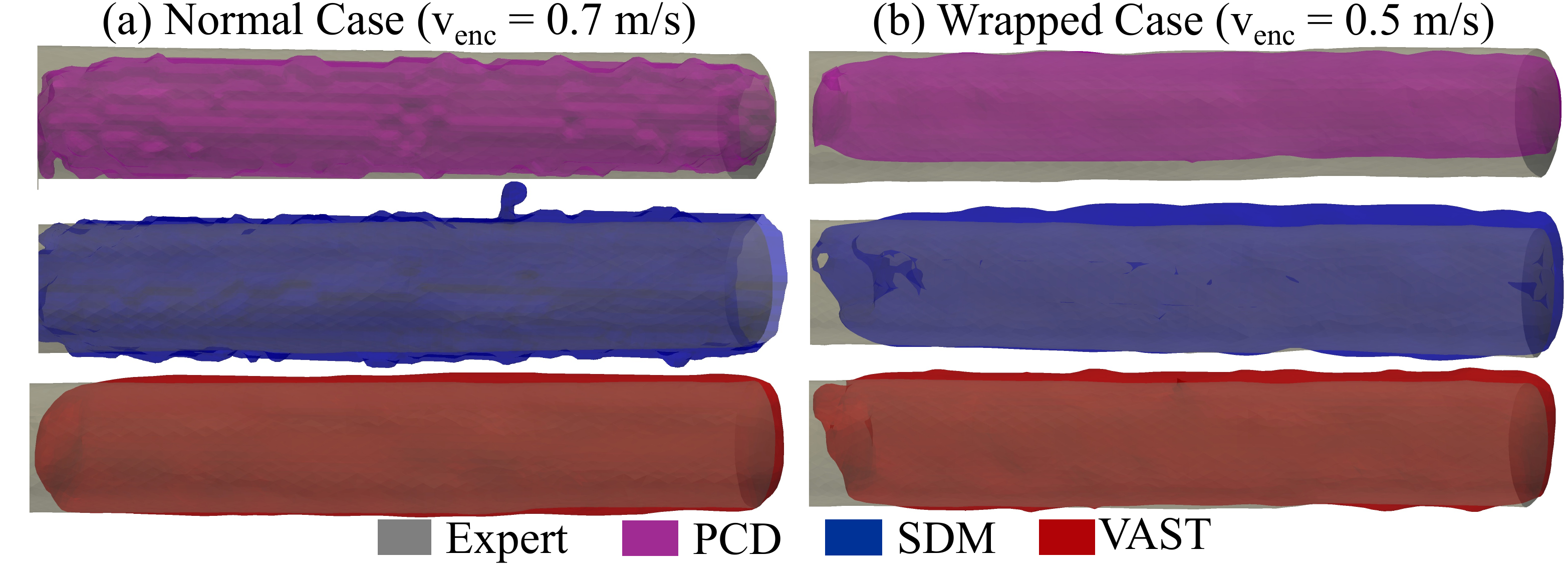}
    \caption{In vitro Poiseuille flow: segmentation surface overlays for the unaliased (\(v_{\text{enc}} = 70~\mathrm{cm/s}\), left) and aliased (\(v_{\text{enc}} = 50~\mathrm{cm/s}\), right) acquisitions. The TOF-derived reference surface of the cylindrical channel is shown together with PCD, SDM, and VAST segmentations to facilitate visual comparison of geometric agreement.}
    \label{fig:invitro_seg_viz}
\end{figure}

\begin{figure}[!t]
    \centering
    \includegraphics[width=\columnwidth]{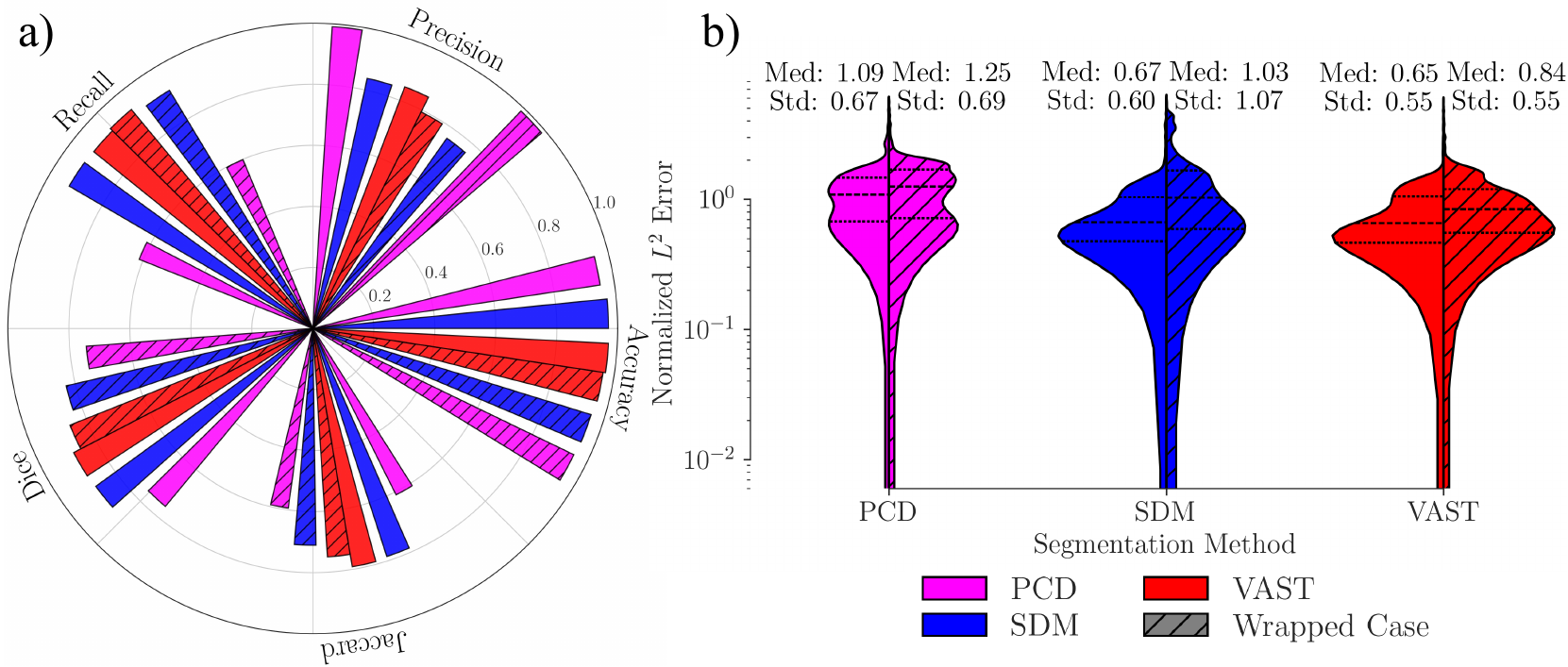}
    \caption{In vitro Poiseuille flow: quantitative segmentation agreement with the TOF reference for PCD, SDM, and VAST under unaliased and aliased acquisitions. (a) Volumetric overlap metrics (definitions in Table~\ref{tab:seg_scores_definition}); hatched bars denote the aliased case. (b) Distributions of normalized surface-to-surface \(L^2\) distances between each automated segmentation and the TOF-derived reference surface for both acquisitions.}
    \label{fig:invitro_seg_scores}
\end{figure}

Segmentation accuracy is assessed by comparing VAST, PCD, and SDM against the TOF-derived reference surface (Figure~\ref{fig:invitro_seg_viz}) and by reporting volumetric overlap and surface-distance metrics (Figure~\ref{fig:invitro_seg_scores}). In this experimental setting, SDM consistently improves upon PCD, which relies on magnitude-driven contrast and therefore tends to under-segment near-wall slow-flow regions. VAST produces the closest agreement with the TOF reference in both acquisitions and remains stable in the presence of background signal impurities that can introduce localized surface roughness in SDM.

Quantitatively, volumetric scores (accuracy, precision, recall, F1-score, Dice, and Jaccard; Table~\ref{tab:seg_scores_definition}) place VAST highest in both the unaliased and aliased scans (Figure~\ref{fig:invitro_seg_scores}a). PCD’s comparatively high precision reflects conservative masks, but its low recall indicates systematic omission of channel voxels, which lowers its F1, Dice, and Jaccard scores. Surface-distance distributions further support these trends: VAST errors concentrate within sub-voxel ranges with a relatively narrow spread across both \(v_{\text{enc}}\) settings, whereas SDM and PCD exhibit broader distributions and heavier tails (Figure~\ref{fig:invitro_seg_scores}b). Overall, the in vitro results support the ranking \(\text{VAST} > \text{SDM} > \text{PCD}\) for segmentation agreement with the TOF reference.

\subsubsection{Flow reconstruction}

\begin{figure}[!t]
    \centering
    \includegraphics[width=\columnwidth]{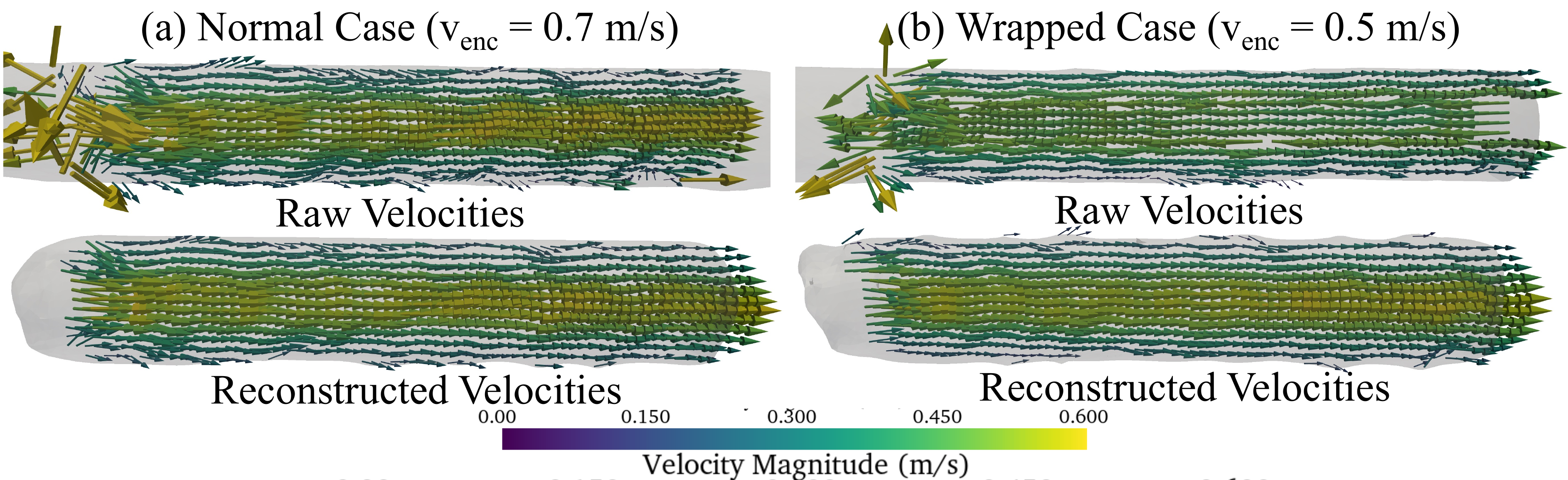}
    \caption{In vitro Poiseuille flow: velocity vectors on a longitudinal plane containing the pipe axis for the unaliased (\(v_{\text{enc}} = 70~\mathrm{cm/s}\), left) and aliased (\(v_{\text{enc}} = 50~\mathrm{cm/s}\), right) acquisitions. Top: raw 4D Flow MRI velocities overlaid with the expert TOF-derived segmentation. Bottom: VAST-reconstructed velocities using the VAST-derived segmentation.}
    \label{fig:invitro_vel_viz}
\end{figure}

We next assess whether VAST recovers the expected Poiseuille velocity profile from the measured 4D Flow MRI data. Figure~\ref{fig:invitro_vel_viz} visualizes velocity vectors on a longitudinal plane containing the pipe axis for both raw and VAST-reconstructed fields. In the unaliased acquisition, the raw field shows localized inlet artifacts and wall-adjacent vectors with non-physical radial components, despite an overall strong signal. In the aliased acquisition, velocities exceeding \(v_{\text{enc}}\) wrap and manifest as direction reversals near the pipe center. VAST suppresses these artifacts by denoising, restoring a coherent axial core, and unwrapping the aliased region, producing a pattern more consistent with laminar pipe flow.

\begin{figure}[!t]
    \centering
    \includegraphics[width=\columnwidth]{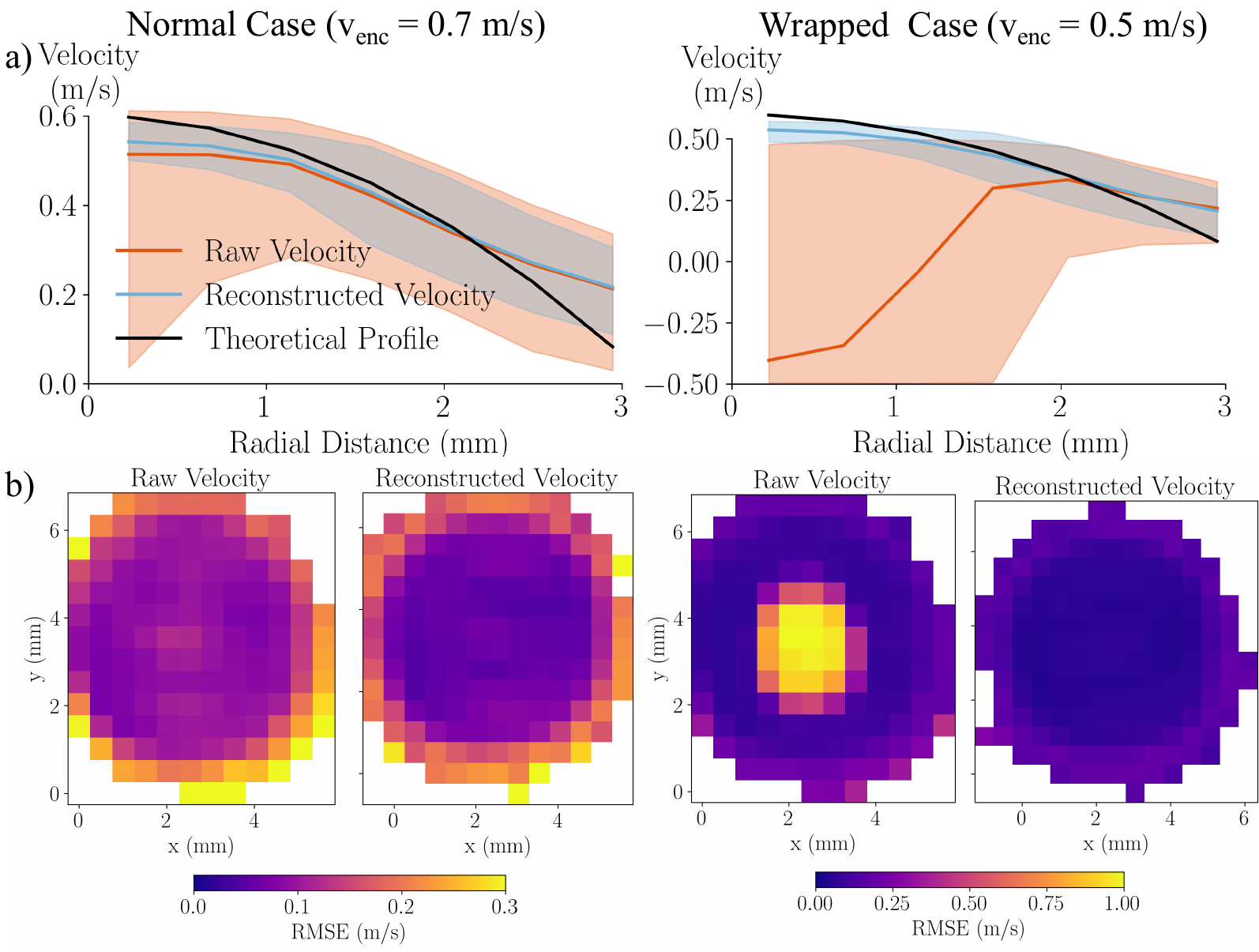}
    \caption{In vitro Poiseuille flow: reconstruction accuracy for the unaliased (\(v_{\text{enc}} = 70~\mathrm{cm/s}\), left) and aliased (\(v_{\text{enc}} = 50~\mathrm{cm/s}\), right) acquisitions. (a) Radially binned velocity profiles for raw (top) and VAST-reconstructed (bottom) fields compared with the analytical parabolic solution. Solid curves denote the median across slices and time frames, and shaded bands indicate the 95\% credible interval; the analytical profile is shown as a single reference curve. (b) Velocity-magnitude error projected onto a representative cross-sectional slice for raw and VAST-reconstructed fields.}
    \label{fig:invitro_vel_score}
\end{figure}

These observations are quantified using radially binned velocity profiles and cross-sectional error maps (Figure~\ref{fig:invitro_vel_score}), together with global RMSE, SSIM, and cosine similarity (Table~\ref{tab:invitro_vel_metrics_comparison}). The radial profiles highlight two dominant failure modes in the raw measurements: near-wall bias (nonzero wall velocity) consistent with segmentation uncertainty and noise, and centerline sign inversions in the aliased scan due to phase wrapping. VAST shifts both profiles toward the analytical parabola, reduces variability across slices and time frames, and removes the aliasing-driven sign error in the wrapped case; a residual near-wall bias remains, which is plausibly attributable to partial-volume effects and residual segmentation mismatch.

The global metrics corroborate these trends (Table~\ref{tab:invitro_vel_metrics_comparison}). In the unaliased scan, VAST reduces RMSE from \(0.174\) to \(0.107~\mathrm{m/s}\) (39\%), increases SSIM from 0.344 to 0.472 (38\%), and increases cosine similarity from 0.896 to 0.955. In the aliased scan, VAST reduces RMSE from \(0.431\) to \(0.100~\mathrm{m/s}\) (77\%), increases SSIM from 0.267 to 0.488 (83\%), and increases cosine similarity from 0.296 to 0.956. Overall, the in vitro experiment indicates that VAST can correct aliasing while reducing noise and preserving the expected laminar flow structure.

\begin{table}[ht]
    \centering
    \caption{Global velocity metrics for raw and VAST-reconstructed velocities in the in vitro Poiseuille flow experiment under unaliased (\(v_{\text{enc}} = 70~\mathrm{cm/s}\)) and aliased (\(v_{\text{enc}} = 50~\mathrm{cm/s}\)) acquisitions. Percentage improvements are reported relative to the raw measurements.}
    \label{tab:invitro_vel_metrics_comparison}
    \renewcommand{\arraystretch}{1.3}
    \small

    \begin{tabular*}{\linewidth}{l@{\extracolsep{\fill}}ccc}
        \hline
        \multicolumn{4}{c}{\textbf{Normal case}} \\
        \hline
        \textbf{Metric} & \textbf{Raw} & \textbf{VAST} & \textbf{Improvement (\%)} \\
        \hline
        RMSE (m/s)        & 0.174 & 0.107 & 38.51 \\
        SSIM              & 0.344 & 0.472 & 37.21 \\
        Cosine similarity & 0.896 & 0.955 & 6.58  \\
        \hline
    \end{tabular*}

    \vspace{1em}

    \begin{tabular*}{\linewidth}{l@{\extracolsep{\fill}}ccc}
        \hline
        \multicolumn{4}{c}{\textbf{Wrapped case}} \\
        \hline
        \textbf{Metric} & \textbf{Raw} & \textbf{VAST} & \textbf{Improvement (\%)} \\
        \hline
        RMSE (m/s)        & 0.431 & 0.100 & 76.79 \\
        SSIM              & 0.267 & 0.488 & 82.88 \\
        Cosine similarity & 0.296 & 0.956 & 223.65 \\
        \hline
    \end{tabular*}
\end{table}

\subsection{In vivo demonstration}\label{sec:invivo}

We next apply VAST to an in vivo intracranial 4D Flow MRI acquisition from the same patient whose internal carotid artery (ICA) aneurysm geometry was used to construct the synthetic benchmark. This test probes whether the pipeline remains stable on clinical data, where reference velocities are unavailable and image quality is constrained by routine acquisition settings.

\begin{figure}[!t]
    \centering
    \includegraphics[width=\columnwidth]{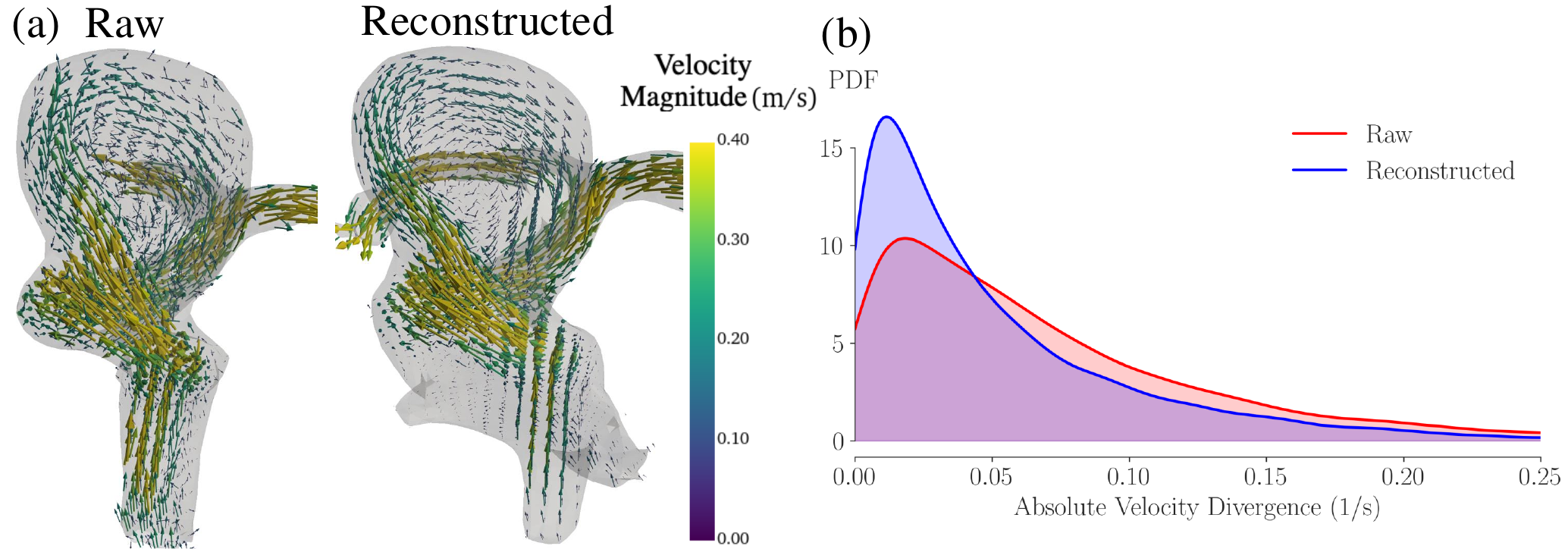}
    \caption{In vivo ICA aneurysm. (a) Peak-systolic velocity vectors overlaid on segmented geometries, comparing raw 4D Flow MRI with expert TOF-based segmentation (left) and VAST-reconstructed velocities with VAST segmentation (right). (b) Empirical probability density functions of voxel-wise divergence residuals for raw and VAST-reconstructed velocity fields, evaluated within the VAST mask.}
    \label{fig:invivo_vel_viz}
\end{figure}

\subsubsection{Segmentation}

\begin{figure}[!t]
    \centering
    \includegraphics[width=\columnwidth]{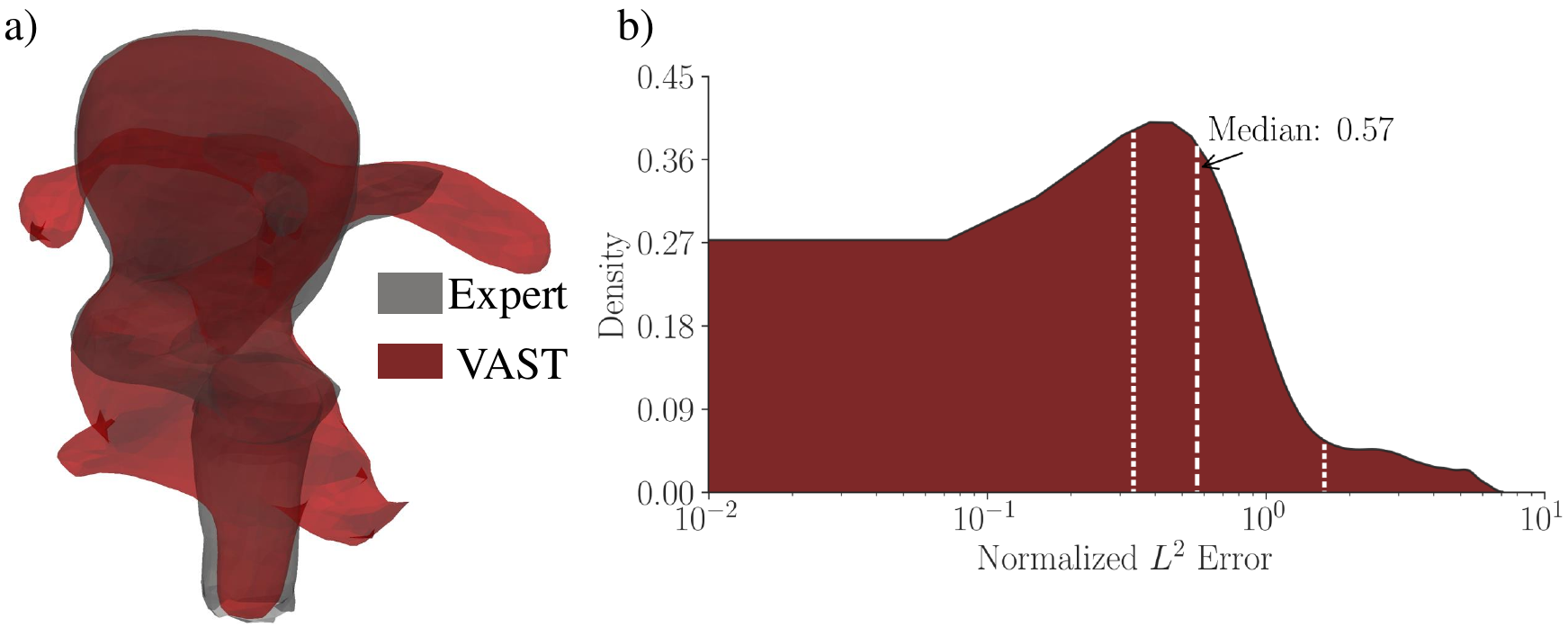}
    \caption{In vivo ICA aneurysm segmentation compared with expert TOF annotation. (a) VAST segmentation (red) overlaid on expert TOF-based segmentation (gray). (b) Distribution of normalized surface-to-surface distances (\(L^2\) errors) between the segmentations; white dotted lines indicate quartiles.}
    \label{fig:invivo_seg_viz}
\end{figure}

VAST delineates the aneurysm dome and parent vessel with close correspondence to the expert TOF-based mask (Figure~\ref{fig:invivo_seg_viz}a). It also includes a small anterior segment that is consistent with a side branch but is not present in the expert annotation. Surface-to-surface distances quantify this agreement: approximately three-quarters of surface points fall within one voxel of the expert mask, and the distribution is concentrated below 57\% of the smallest voxel dimension (\(1.1 \times 1.1 \times 1.3~\text{mm}^3\); Figure~\ref{fig:invivo_seg_viz}b).

Volumetric overlap metrics are correspondingly high (accuracy 0.98, precision 0.73, recall 0.86, Dice 0.79, Jaccard 0.65). The lower Jaccard index (0.65) is consistent with VAST including additional connected vessel segments beyond the expert-defined ICA region; the current implementation produces a single vascular mask rather than an anatomy-specific or multi-label segmentation. Enabling vessel-specific labeling is a natural extension for improving overlap-based scores without changing the underlying voxel-wise detection strategy.

\subsubsection{Flow reconstruction}

Peak-systolic vector fields (Figure~\ref{fig:invivo_vel_viz}a) show that the raw measurement is broadly interpretable but contains localized irregular vectors near the wall and within the recirculation region, where velocities approach the noise floor and are most susceptible to phase errors. VAST reduces these localized inconsistencies and yields a more coherent representation of the jet and intra-aneurysmal circulation while remaining visually consistent with the measured flow patterns.

Without an in vivo velocity ground truth, we evaluate reconstruction quality using divergence residuals as an internal consistency check (Figure~\ref{fig:invivo_vel_viz}b). The VAST-reconstructed field shifts the residual distribution toward zero and suppresses the tail of large residuals, consistent with improved adherence to incompressibility within the segmented domain. Quantitatively, the mean residual decreases by 30.6\% (0.073 to 0.051) and the interquartile range decreases by 27.1\% (0.077 to 0.056), indicating improved mass-consistent flow while remaining compatible with the acquired data.

\subsection{Computational performance}\label{sec:computational}

We characterize runtime and convergence to assess practicality. Across all datasets, segmentation converged within five iterations per cardiac phase. Reconstruction converged in five iterations for the lowest-SNR synthetic case and in two iterations for the in vitro and in vivo datasets.

Table~\ref{tab:computational_times} reports runtimes and grid sizes on a 14-core Apple M3 Pro CPU (no GPU accelerations). End-to-end processing completed in minutes: the largest volume (in vitro) finished in \(\sim 200\)~s, while the synthetic and in vivo aneurysm cases finished in \(\sim 90\)~s, supporting deployment on standard workstation hardware.

\begin{table}[htbp]
    \centering
    \caption{Computational runtimes for segmentation and reconstruction across synthetic, in vitro, and in vivo datasets, including dataset dimensions and number of reconstruction iterations, computed on a 14-core Apple M3 Pro CPU.}
    \label{tab:computational_times}
    \resizebox{\columnwidth}{!}{%
    \begin{tabular}{lccccccc}
        \toprule
        \textbf{Dataset} & $N_t$ & $N_x$ & $N_y$ & $N_z$ & \textbf{Segmentation (s)} & \textbf{Reconstruction (s)} & \textbf{Iterations} \\
        \midrule
        Synthetic (SNR = 2)  & 13  & 20  & 21  & 24  & 17.65  & 61.00  & 5  \\
        In vitro             & 10  & 30  & 30  & 71  & 70.62  & 129.15 & 2  \\
        In vivo              & 13  & 30  & 25  & 30  & 34.22  & 46.18  & 2  \\
        \bottomrule
    \end{tabular}%
    }
\end{table}

\section{Conclusion}\label{sec:conclusion}

We presented VAST, a fully automated workflow for intracranial 4D Flow MRI that integrates unsupervised vessel segmentation with physics-informed velocity reconstruction. Across synthetic aneurysm benchmarks spanning low SNR and strong phase wrapping, VAST maintains sub-voxel geometric agreement and substantially improves velocity agreement with CFD references (lower RMSE, higher SSIM/cosine similarity) relative to raw measurements. In vitro, it closely matches TOF-derived channel geometry and recovers velocity profiles consistent with analytical Poiseuille flow, including in aliased acquisitions. In vivo, VAST closely matches expert TOF masks and reduces divergence residuals by \(\sim 30\%\), indicating improved physical consistency.

VAST completes in minutes on a standard CPU with no user interaction, enabling reproducible processing without manual segmentation or acquisition-specific tuning. By coupling adaptive background modeling with continuity-aware reconstruction, VAST provides a practical path toward more reliable intracranial 4D Flow quantification under routine acquisition constraints. Future work will extend VAST toward anatomy-specific and multi-label segmentation, incorporate explicit correction of background phase offsets (e.g., eddy-current and field-inhomogeneity effects), and support direct estimation of clinically relevant hemodynamic biomarkers such as wall shear stress and relative pressure across larger cerebrovascular territories and cohorts.

\section*{Acknowledgment}

The authors thank Neal Minesh Patel for assisting with the in vitro Poiseuille flow experiment and segmenting flow from time-of-flight images. They also thank Rudra Sethu Viji for preparing the PDMS block and the blood analog solution, and for supporting the in vitro flow experiment. Additionally, the authors gratefully acknowledge Professor Michael Markl (Northwestern University) for providing the in vivo 4D Flow and time-of-flight angiography data.

\section*{Funding}

This work was supported by the National Institutes of Health (NIH), National Heart, Lung, and Blood Institute (NHLBI) under grant R01 HL115267.

% ---- Appendices (included in the main manuscript; no separate supplementary file) ----
% -----------------------------------------------------------------------------
% Appendices (kept concise to meet IEEE TMI length constraints)
% -----------------------------------------------------------------------------
\appendices

\section{Implementation details of VAST}
\label{app:technical}

\subsection{Segmentation}

\subsubsection{Magnitude denoising via tensor decomposition}
\label{app:magnitude_denoising}

We represent magnitude images as a 4D tensor \(\mathrm{Mag}(x,y,z,t)\) and apply Tucker decomposition using \texttt{TensorLy}~\cite{tensorly}:
\begin{equation}
\mathrm{Mag}(\mathbf{x}, t) \approx \sum_{i=1}^{r_x} \sum_{j=1}^{r_y} \sum_{k=1}^{r_z} \sum_{m=1}^{r_t} 
\lambda_{ijkm}\, S_i(x)\, S_j(y)\, S_k(z)\, T_m(t),
\end{equation}
where \(S_i,S_j,S_k\) are spatial factors, \(T_m\) temporal factors, and \(\lambda_{ijkm}\) core coefficients. To suppress noise while preserving gross temporal structure, we retain only the dominant temporal mode \(m=1\):
\begin{equation}
\mathrm{Mag}_1(\mathbf{x}, t) = \sum_{i=1}^{r_x} \sum_{j=1}^{r_y} \sum_{k=1}^{r_z} \lambda_{ijk1}\, S_i(x)\, S_j(y)\, S_k(z)\, T_1(t),
\end{equation}
and use \(\mathrm{Mag}_1\) for segmentation.

\subsubsection{Initial background classification}
\label{app:initial_background}

We normalize \(\mathrm{Mag}_1\) to \([0,1]\) and apply 3D Sauvola thresholding~\cite{sauvola} (\texttt{scikit-image}~\cite{scikit-image}). The local threshold is
\begin{equation}
\mathrm{thresh}^{(1)}(\mathbf{x},t) =
m(\mathbf{x},t)\left[1 + k\left(\frac{\sigma(\mathbf{x},t)}{R} - 1\right)\right],
\end{equation}
computed over an \(11\times 11\times 11\) neighborhood, with \(k=0.2\) and \(R=0.5\). Voxels with
\(\mathrm{Mag}_1(\mathbf{x}, t) < \mathrm{thresh}^{(1)}(\mathbf{x}, t)\) are labeled background and the complement defines the initial flow mask.

\subsubsection{Magnitude-based background likelihood}
\label{app:mag_likelihood}

At iteration \(n\), we form the sample set of background magnitudes
\(\{s_i^{(n)}\}=\{\mathrm{Mag}_1(\mathbf{x},t)\mid \mathrm{Mask}_{\mathrm{bg}}^{(n-1)}(\mathbf{x},t)=1\}\)
and fit a smooth background density using SciPy’s Gaussian KDE~\cite{scipy}. Let \(F_{\mathrm{bg}}^{(n)}\) denote the resulting CDF. The magnitude-based background likelihood is defined as the upper-tail probability
\begin{equation}
p_{\mathrm{mag}}^{(n)}(\mathbf{x},t)
= 1 - F_{\mathrm{bg}}^{(n)}\bigl(\mathrm{Mag}_1(\mathbf{x},t)\bigr),
\end{equation}
so unusually bright voxels under the background model receive low background likelihood.

\subsubsection{Phase-based likelihood (SDM extension)}
\label{app:phase_likelihood}

We extend SDM~\cite{sdm} using a \(3\times3\times3\times3\) spatiotemporal neighborhood (spatial stencil \(\mathcal{N}(\mathbf{x})\) and three time frames \(t-1,t,t+1\)). For each velocity component \(U_i(\mathbf{x},t)\),
\begin{align}
\mathbb{E}_{\mathbf{x},t}[U_i] &= \frac{1}{N}\sum_{\tau=t-1}^{t+1}\sum_{\mathbf{x}'\in\mathcal{N}(\mathbf{x})} U_i(\mathbf{x}',\tau),\\[2pt]
\sigma_{\mathbf{x},t}[U_i] &= 
\sqrt{\frac{1}{N}\sum_{\tau=t-1}^{t+1}\sum_{\mathbf{x}'\in\mathcal{N}(\mathbf{x})}
\bigl(U_i(\mathbf{x}',\tau) - \mathbb{E}_{\mathbf{x},t}[U_i]\bigr)^2},
\end{align}
with \(N\) the number of samples in the neighborhood. Assuming background phase is zero-mean noise, we define
\begin{equation}
\tilde{U}_i(\mathbf{x},t)
= \frac{\mathbb{E}_{\mathbf{x},t}[U_i]}{\sigma_{\mathbf{x},t}[U_i]/\sqrt{3}},
\qquad
\tilde{U}(\mathbf{x},t)=\sqrt{\sum_{i=1}^3\tilde{U}_i^2(\mathbf{x},t)}.
\end{equation}
We convert \(\tilde{U}\) to a phase-based background likelihood using SDM calibration recomputed as the mask updates:
\begin{equation}
p_{\mathrm{phase}}^{(n)}(\mathbf{x},t)
=\mathrm{SDMcal}\bigl(\tilde{U}(\mathbf{x},t);\mathrm{Mask}^{(n-1)}(\mathbf{x},t)\bigr).
\end{equation}

\subsubsection{Likelihood fusion via total-variation regularization}
\label{app:likelihood_fusion}

We fuse likelihoods in log space:
\begin{equation}
\begin{aligned}
\log p_{\text{comb}}^{(n)}(\mathbf{x}, t; w)
&= w(\mathbf{x}, t)\,\log p_{\text{mag}}^{(n)}(\mathbf{x}, t) \\
&\quad + \bigl(1 - w(\mathbf{x}, t)\bigr)\,\log p_{\text{phase}}^{(n)}(\mathbf{x}, t) .
\end{aligned}
\end{equation}
with \(0\leq w\leq 1\). The weight field is selected via TV regularization:
\begin{equation}
\min_{0\le w\le 1}\ \mathrm{TV}\!\left(\log p_{\text{comb}}^{(n)}(\cdot,\cdot; w)\right),
\end{equation}
implemented in \texttt{cvxpy}~\cite{cvxpy}. We min--max normalize \(\log p_{\text{comb}}^{(n)}\) to \([0,1]\), apply Sauvola thresholding to obtain the updated background mask, define \(\mathrm{Mask}^{(n)}=1-\mathrm{Mask}_{\mathrm{bg}}^{(n)}\), and perform vessel isolation/hole filling as in SDM~\cite{sdm}. Iteration stops when adaptive thresholds stabilize within 1\%.

\subsection{Velocity reconstruction}

\subsubsection{Continuity-constrained phase unwrapping}
\label{app:unwrapping}

Let \(\psi\) be wrapped phase and \(\phi\) unwrapped phase, with \(\psi=\mathrm{wrap}(\phi)\in[-\pi,\pi]\). We construct wrapped gradient observations \(\widehat{\nabla_r\phi}\) and solve our continuity-constrained unwrapping problem~\cite{unwrapping}:
\begin{equation}
\begin{aligned}
\hat\phi &=
\arg\min_{\phi}\;
\Bigl\|W\bigl(D_r \phi - \widehat{\nabla_r\phi}\bigr)\Bigr\|_2^2 \\
&\quad + \Bigl\|\frac{\mathrm{venc}_u}{\pi}D_x\phi_u
+ \frac{\mathrm{venc}_v}{\pi}D_y\phi_v
+ \frac{\mathrm{venc}_w}{\pi}D_z\phi_w\Bigr\|_2^2
+ \alpha\|\phi\|_2^2 .
\end{aligned}
\end{equation}
where \(D_r\) are finite differences, \(\alpha=0.01\), and
\begin{equation}
W = \mathrm{diag}\!\left(\frac{1}{\sigma_{\widehat{\nabla_r\phi}}^{2}}\right).
\end{equation}
We solve using ridge regression (\texttt{scikit-learn}~\cite{scikit-learn}) with Dirichlet boundary conditions set by the median boundary phase.

\subsubsection{Universal outlier detection}
\label{app:uod}

For each velocity component \(U_i(\mathbf{x})\), we compute the normalized residual (UOD~\cite{uod})
\begin{equation}
R_{\mathrm{norm}}(\mathbf{x})
= \frac{\bigl|U_i(\mathbf{x}) - \mathrm{median}\bigl(\mathcal{N}(\mathbf{x})\setminus\{\mathbf{x}\}\bigr)\bigr|}
       {\mathrm{median}\bigl\{\bigl|U_i(\mathbf{x}') - \mathrm{median}(\mathcal{N}(\mathbf{x}')\setminus\{\mathbf{x}'\})\bigr|\bigr\} + \epsilon},
\end{equation}
with \(\mathcal{N}(\mathbf{x})\) a \(3\times3\times3\) neighborhood, \(\mathbf{x}'\in\mathcal{N}(\mathbf{x})\), and \(\epsilon=10^{-3}\).
A voxel is marked as an outlier if \(R_{\mathrm{norm}}(\mathbf{x})>\tau\) with \(\tau=2\); flagged values are replaced by the neighborhood median.

\subsubsection{POD-based velocity denoising and convergence}
\label{app:pod}

We apply POD to the stacked velocity field \(\mathbf{U}(\mathbf{x},t)\):
\begin{equation}
\mathbf{U}(\mathbf{x}, t)
= \sum_{k=1}^{N} a_{k}(t)\,\Phi_{k}(\mathbf{x}).
\end{equation}
Mode entropies are computed from a spatial DCT and clustered with DBSCAN to identify informative modes \(\mathcal{K}\)~\cite{dbscan, scikit-learn}. The denoised field is
\begin{equation}
\mathbf{U}_{\mathrm{filtered}}(\mathbf{x}, t)
= \sum_{k\in\mathcal{K}} a_{k}(t)\,\Phi_{k}(\mathbf{x}).
\end{equation}

\subsubsection{Iterative reconstruction and convergence}
\label{app:iteration}

Phase unwrapping, UOD, and POD denoising are repeated in an outer loop. At iteration \(i\), we compute the signal energy
\begin{equation}
E^{(i)} = \sum_{k\in\mathcal{K}} D_k^{(i)},
\end{equation}
where \(D_k^{(i)}\) are the POD eigenvalues. We monitor the relative change
\begin{equation}
\frac{|E^{(i)}-E^{(i-1)}|}{E^{(i-1)}}.
\end{equation}
The algorithm terminates when this ratio falls below \(0.01\), and we roll back to iteration \(i-1\) and use that reconstruction as the final velocity field.

\section{Synthetic 4D Flow MRI generation}
\label{app:synthetic}

We generate synthetic 4D Flow MRI from a transient CFD simulation of a patient-specific ICA aneurysm as in~\cite{brindise_multimodality}. The CFD velocity field \(\mathbf{U}(\mathbf{r}_{\mathrm{CFD}},t)\) is linearly interpolated onto the 4D Flow voxel grid \(\mathbf{x}\) (voxel sizes \(\Delta x,\Delta y,\Delta z\)), yielding component fields \(U_i(\mathbf{x},t)\), \(i\in\{1,2,3\}\). We then synthesize complex phase-contrast signals for each encoding direction:
\begin{equation}
M_i(\mathbf{x},t)=\mathrm{Mag}_i(\mathbf{x},t)\exp\!\left(\frac{j\pi\,U_i(\mathbf{x},t)}{\mathrm{venc}}\right),
\label{eq:synthetic_signal_app}
\end{equation}
with a prescribed magnitude model \(\mathrm{Mag}_i=1\) inside the lumen and \(\mathrm{Mag}_i=0.2\) in background (reference channel assigned zero phase). Wrapped phases are obtained as \(\psi_i(\mathbf{x},t)=\arg(M_i(\mathbf{x},t))\in[-\pi,\pi]\).

To approximate intravoxel averaging from finite k-space sampling, each complex channel is convolved with a truncated 3D sinc point-spread function~\cite{rispoli_synMRI}:
\begin{equation}
M_{i,\mathrm{blur}}(\mathbf{x},t)=[M_i(\cdot,t)\ast K](\mathbf{x}),\quad
\end{equation}
\begin{equation}
    K(x,y,z)=\mathrm{sinc}\!\Bigl(\frac{x}{\Delta x}\Bigr)\mathrm{sinc}\!\Bigl(\frac{y}{\Delta y}\Bigr)\mathrm{sinc}\!\Bigl(\frac{z}{\Delta z}\Bigr).
\end{equation}
with truncation \(K=0\) for \(|x|>2\Delta x\), \(|y|>2\Delta y\), or \(|z|>2\Delta z\). Complex Gaussian noise is then added independently per channel,
\begin{equation}
\tilde{M}_i(\mathbf{x},t)=M_{i,\mathrm{blur}}(\mathbf{x},t)+n_i(\mathbf{x},t),\qquad
n_i\sim\mathcal{CN}(0,\sigma^2),
\end{equation}
where the noise level is set to achieve a target SNR via
\begin{equation}
\sigma(\mathbf{x},t)=\mathrm{Mag}_i(\mathbf{x},t)/\mathrm{SNR}.
\label{eq:snr_def_app}
\end{equation}

We generate an SNR sweep \(\mathrm{SNR}\in\{20,10,5,3,2\}\). To probe aliasing, we fix \(\mathrm{SNR}=10\) and sweep \(\mathrm{venc}\in\{1.0,0.5,0.4,0.3,0.2\}\,v_{\max}\), where \(v_{\max}=\max_{\mathbf{x},t}\|\mathbf{U}(\mathbf{x},t)\|\); this yields wrapping factors up to \(\sim 5\times\).

\bibliographystyle{IEEEtran}
\bibliography{bibliography}

\end{document}